\documentclass[aps,prl,twocolumn,groupedaddress,showpacs]{revtex4}

\newcommand{\rbf}       {\mbox{\boldmath$r$}}
\newcommand{\vmbf}       {\mbox{\boldmath$v_m$}}
\newcommand{\vsbf}       {\mbox{\boldmath$v_s$}}
\newcommand{\fbf}       {\mbox{\boldmath$f_0$}}
\newcommand{\fbfvn}       {\mbox{\boldmath$f_{vn}$}}
\newcommand{\ubf}       {\mbox{\boldmath$u$}}
\newcommand{\vbf}       {\mbox{\boldmath$v$}}
\newcommand{\kappabf}       {\mbox{\boldmath$\kappa$}}
\newcommand{\Fmfbf}       {\mbox{\boldmath$F_{mf}$}}
\newcommand{\nubf}       {\mbox{\boldmath$\nu$}}
\newcommand{\omegabf}       {\mbox{\boldmath$\omega$}}
\newcommand{\fdbf}       {\mbox{\boldmath$f_d$}}

\def\eg{{\it e.g.}}

\def\gap{\hbox{${_{\displaystyle>}\atop^{\displaystyle\sim}}$}}

\usepackage{graphicx}
\usepackage{subfigure}

\begin{document}

\voffset=1.cm


\title{Dynamics of Quantum Vorticity in a Random Potential} 


\author{Bennett Link}
\affiliation{Montana State University, Department of Physics, Bozeman MT
59717}
\email[]{link@physics.montana.edu}
\thanks{}


\date{\today}

\begin{abstract}
I study the dynamics of a superfluid vortex in a random potential, as
in the inner crust of a neutron star. Below a critical flow velocity
of the ambient superfluid, a vortex is effectively immobilized by
lattice forces even in the limit of zero dissipation. Low-velocity,
translatory motion is not dynamically possible, a result with
important implications for understanding neutron star precession and
the dynamical properties of superfluid nuclear matter.
\bigskip\bigskip
\end{abstract}

\pacs{97.60.Jd, 26.60.-c, 47.37.+q, 97.60.Gb}

\bigskip

\maketitle


A neutron star (NS) is expected to comprise over
a solar mass of distinct quantum liquids. In the inner crust (between
the drip density and approximately half nuclear density), a $^1_0S$
neutron superfluid (SF) threaded by an array of quantized vortices
coexists with the ionic lattice. Forces exerted by the vortex array on
the lattice could have important observable effects on NS spin and
thermal evolution (\eg,
\cite{shaham77,alpar_etal84,sl89,le96,swc99}), especially 
if vortices pin to the lattice as suggested long ago \cite{ai75}.
Evidence for long-period precession in some NSs (\eg, \cite{sls00}),
an interpretation supported by quantitative modeling \cite{alw06}, is
difficult to explain if vortices pin in the inner crust
\cite{shaham77} and presents challenges to 
the standard picture of the outer core
\cite{link03}. The critical question of whether or not pinning occurs in the
inner crust has been studied (\eg, \cite{jones97,jones98}) but not yet
satisfactorily answered. Here I show that a vortex is trapped by the
inner-crust lattice under very general conditions if the ambient SF
velocity is below a critical value.  The problem of vortex dynamics
and pinning in a lattice potential is also of considerable interest in
laboratory Bose-Einstein condensates (see,
\eg, \cite{bhc06}). 
 
{\em The vortex drag description.}--Most hydrodynamic studies of the
coupling problem begin by including a term in the SF acceleration
equation for the ``mutual friction'' force in a homogeneous medium.
For a SF of mass density $\rho_s$ flowing at velocity $\vsbf$, coupled
to a medium moving at velocity $\vmbf$, and neglecting the collective
tension of the vortex lattice, the mutual friction force density in
the non-rotating frame is \cite{hv56,bk61}
\begin{equation}
\Fmfbf=\beta^\prime\rho_s\omegabf\times(\vsbf-\vmbf)+
\beta\rho_s\nubf\times\left[\omegabf\times(\vsbf-\vmbf)\right],
\label{mf}
\end{equation}
where $\omegabf\equiv\nabla\times\vsbf$ is the SF vorticity, $\nubf$
is a unit vector in the direction of $\omegabf$, and $\beta$ and
$\beta^\prime$ are coefficients to be determined by a microscopic
calculation for the application of interest. This drag term, or
extensions of it to bulk neutron matter in beta equilibrium, has
been widely used in studies of NS hydrodynamics and precession; the
``medium'' could be taken to be, for example, the lattice of the NS
inner crust, magnetic flux tubes of the outer core with which the vortices
interact, and the charged fluid of the outer core. The mutual friction
force of eq. [\ref{mf}] is directly related to the drag force per unit
length on a vortex \cite{swc99}
\begin{equation}
\fdbf=-\eta\vbf-\eta^\prime\nubf\times\vbf,
\label{fd}
\end{equation}
where $\vbf$ is the vortex velocity with respect to the medium, and
the coefficients $\eta$ and $\eta^\prime$ are related to $\beta$ and
$\beta^\prime$. The second term in eq. [\ref{fd}] is non-dissipative
and is usually assumed to be zero, as I also assume; in this case
$\beta=\eta_r/(1+\eta_r^2)$ and $\beta^\prime=\eta_r\beta$, where
$\eta_r\equiv\eta/\rho_s\kappa$ is the reduced drag coefficient and
$\kappa$ is the vorticity quantum $h/2m_n$ ($m_n$ is the neutron mass
for a neutron SF). Dissipative processes that act on a translating
vortex include electron scattering \citep{feibelman71,be89}, and the
excitation of vortex waves (Kelvin modes) through the vortex-nucleus
interaction
\citep{eb92,jones92}. All of these calculations give $\eta_r<<1$. 

Neglecting vortex bending and local non-dissipative forces, the motion
of a vortex segment follows from equating the sum of the Magnus and
drag forces to zero: 
\begin{equation}
\rho_s\kappabf\times\left(\vbf-\vsbf\right)
-\eta \vbf = 0, 
\label{drag}
\end{equation}
where $\kappabf$ is aligned with the vortex (the $z$ axis) and $\vsbf$
is now the velocity of the ambient SF in the rest frame of the medium,
taken to be along the $-x$ axis. In this description (henceforth, the
``drag description''), the vortex segment moves at velocity
$\vbf\simeq\vsbf$ for low drag ($\eta_r<<1$) and $\vbf\simeq 0$ for
high drag ($\eta_r>>1$). The large drag limit has been taken as
corresponding to the pinning of vortices to defects, such as nuclei in the
inner crust or magnetic flux tubes in the outer core
\cite{swc99,sedrakian05,link06,gaj08a}. 
The drag description embodied in eq. [\ref{drag}], however,
incorrectly describes the motion of a vortex through these
fundamentally {\em inhomogeneous} environments. A correct description
of vortex motion in the inner crust requires the inclusion of two
additional forces: 1) the local, {\em non-dissipative} component of
the force exerted on the vortex by the lattice, and, 2) the elastic
force of the vortex. These forces qualitatively change the vortex
motion. Appropriate to the inner crust (but not the core), I focus
on a single-component SF. 

{\em Vortex motion through a random lattice.}--Consider the motion of
a vortex that is dissipatively coupled to a background lattice. The 
displacement vector of the vortex with respect to the $z$-axis is
$\ubf(z,t)=u_x(z,t)\hat{x}+u_y(z,t)\hat{y}$. The equations of motion
for a vortex moving in the absence of external forces can be
found in Sonin \cite{sonin}; a hydrodynamic description suffices
for excitation wavelengths significantly larger than the SF coherence
length $\xi$, about 10 fm in the inner crust. The force per unit
length exerted on the vortex by the lattice has a non-dissipative
contribution $\fbf$ and a dissipative contribution taken here to be
the drag force of eq. [\ref{fd}] with $\eta^\prime=0$, assumed to
hold locally, and approximated as linear in the local vortex velocity
(but easily generalized). The equations of motion become
\begin{equation}
T_v\frac{\partial^2\ubf}{\partial z^2}+
\rho_s\kappabf\times\left(\frac{\partial\ubf}{\partial t}-\vsbf\right)
+ \fbf -\eta \frac{\partial\ubf}{\partial t} = 0, 
\label{eomv}
\end{equation}
where $T_v=(\rho_s\kappa^2/4\pi)\ln (\xi k)^{-1}$ is the vortex
self-energy (tension) for an excitation of wave number $k$. The first
term represents the restoring force of tension as the vortex is
bent. The second term is the Magnus force per unit length exerted on a vortex
that is moving with respect to the ambient SF. In the absence of
external forces, the vortex would remain straight and motionless with
respect to the solid. An essential feature of vortex dynamics is that
the local vortex velocity is determined entirely by external forces
and by the shape of the vortex, and not by inertial forces (for
temperatures small compared to the pairing gap). In
the absence of external forces, the solutions to eqs. [\ref{eomv}] are
circularly-polarized, diffusive waves (Kelvin waves) with frequencies
$\omega_k = \pm T_v k^2/\rho_s\kappa$ \cite{kelvin}. Characteristic
bending scales of the vortex in the problem are $\sim 500$ fm (see
below), and fixing $\ln (\xi k)^{-1}=3$ is an adequate approximation. I
ignore quantum effects on the vortex motion and thermal excitations.

Nuclei in the solid with which the superfluid coexists exert local 
forces on the vortex which vary over a length scale of order the
average nuclear spacing $a$, typically 30-50 fm in the inner
crust. Calculations of the vortex-nucleus interaction energy $E_{vn}$
give values of up to $\sim 1$ MeV, attractive
in some regions and repulsive in others
\citep{dp06,abbv07}; the corresponding characteristic force on a vortex
segment of length $a$ is $F_m\equiv a^{-1}E_{vn}$.  
The solid is most likely amorphous
\cite{jones01}, in which case the vortex interacts with an 
effectively {\em random potential} 
Even if the lattice does possess long-range crystalline
order, vortices will be aligned with the SF rotation axis on average
which will not, in general, coincide with any symmetry axis of the
solid; a random lattice approximation should be adequate also in this
situation. In the denser regions of the NS inner crust, $T_v$ is of
order an MeV fm$^{-1}$. Defining a dimensionless strength parameter
$s\equiv F_m/T_v$ ($>0$ for an attractive potential), estimates of
$F_m$ from Refs. \cite{dp06,abbv07} give $10^{-2}\le\vert s\vert\le
0.1$. A vortex is thus a stiff object; the vortex-nucleus interaction
can bend a vortex by an amount $a$ only over a length scale of order
$\gap 10a$
\cite{lc02}; a vortex segment of length $a$ can be regarded as
straight to a good approximation and the lattice force on the vortex
can be taken as acting in the $x-y$ plane at any given $z$ along the
vortex. For the force exerted by a single nucleus at the origin on a
vortex segment of length $a$, I take a parameterized central force in
the $x-y$ plane:
\begin{equation}
\fbfvn(\ubf)=-1.7F_m\frac{\ubf}{r_p}{\rm e}^{-u^2/2r_p^2} 
\label{force}
\end{equation}
The force has a maximum magnitude of $F_m$ at $u=r_p$, where $r_p$ is
the effective range of the potential, henceforth taken to be
$r_p=a/2$.  Dividing the vortex into segments of length $a$, let the
total non-dissipative force on the $i$-th segment be
\begin{equation}
\fbf(z_i) =  \sum_j \fbfvn(\ubf_i-\rbf_j),
\label{fl}
\end{equation}
where the nuclei are randomly placed at locations $\rbf_j$ in planes
separated by $a$ and parallel to the $x-y$ plane. For simplicity, each
summation is only over nuclei in the $i$-th plane.  In the following,
lengths will be expressed in units of $a$, time in units of $t_D\equiv
\rho_s\kappa a^2/T_v$ (the characteristic diffusion time of a Kelvin
wave of wavenumber $k=a^{-1}$ along the segment) and velocities in
units of $a/t_D$. As described below, the sign of $s$ is unimportant
for $r_p\sim a$; I choose $s>0$, corresponding to attractive nuclear
potentials, for illustration. I focus on the regime
$0.1\ge\eta_r\ge 0$ which widely brackets the range found by all drag
calculations, and take $s=0.1$. 

\begin{figure}[t]
\centering
\subfigure
{
    \includegraphics[width=.45\linewidth]{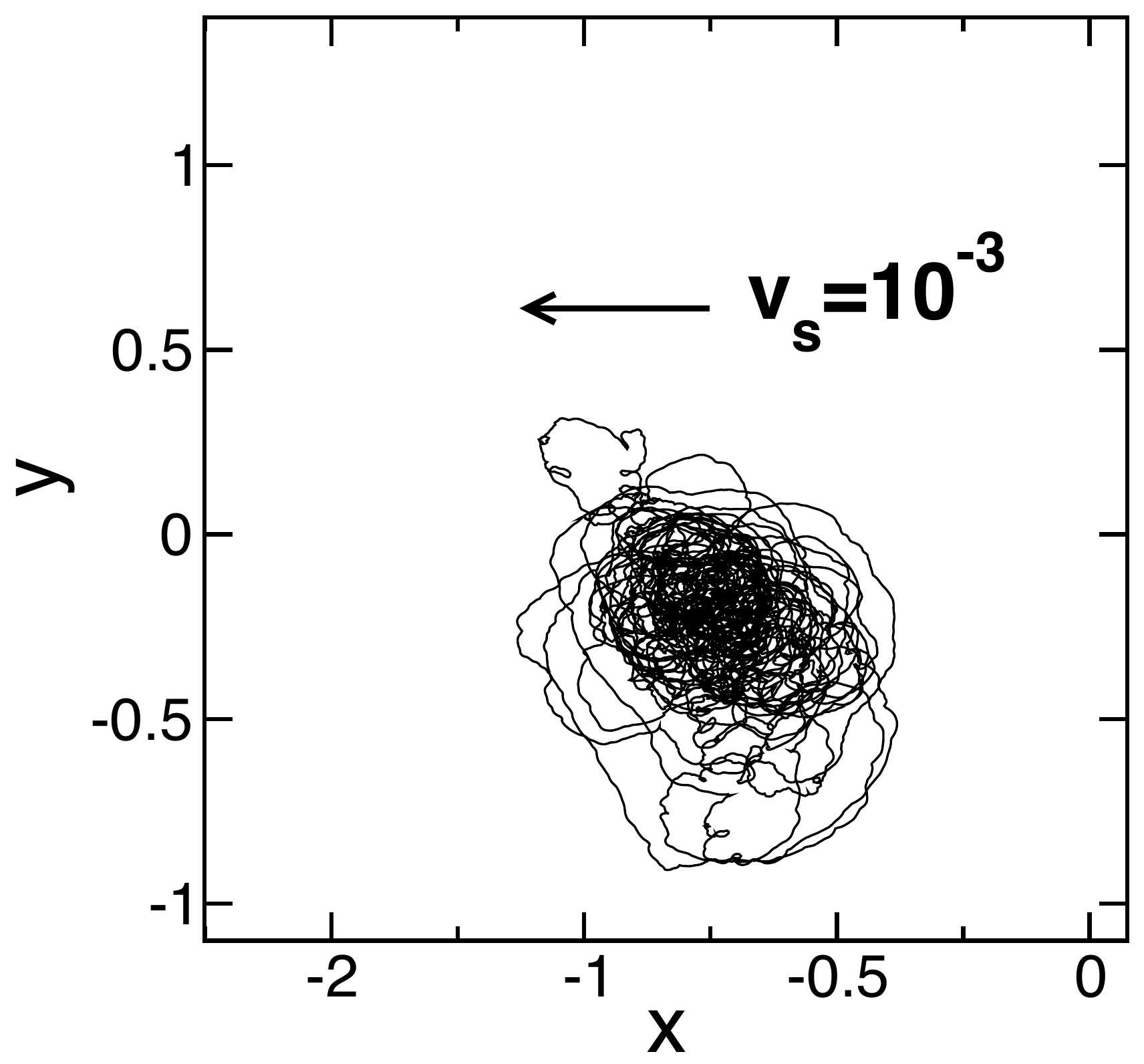}
}
\subfigure
{
     \includegraphics[width=.47\linewidth]{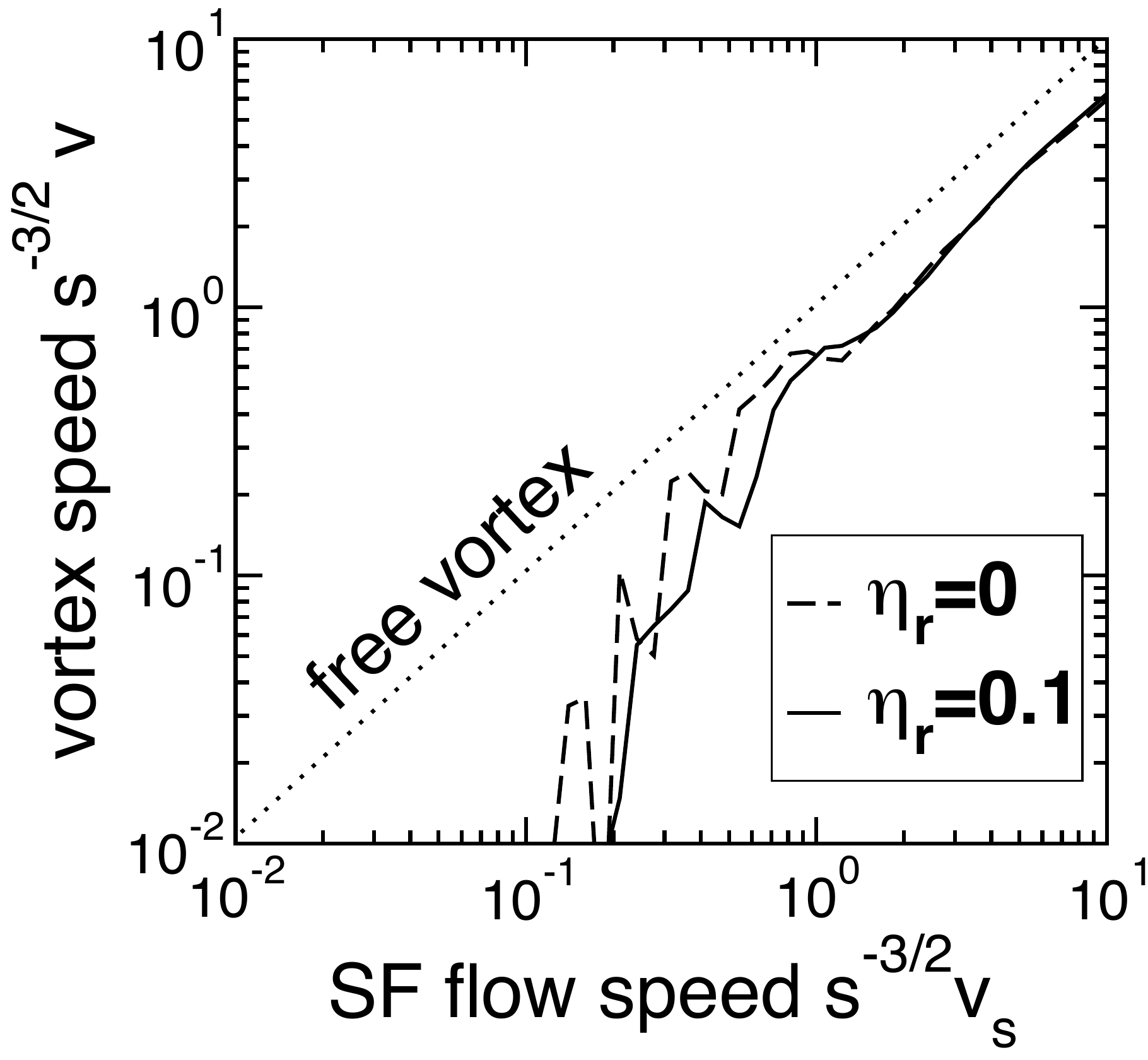}
}
\caption{{\em Left panel:} vortex position, averaged over its
length, in the presence of an external SF flow field of velocity
$v_s=10^{-3}$ and no dissipation over $5\times 10^3$ time
units. Without the random potential, the vortex would translate to the
left with the SF by five units. The non-dissipative lattice force
effectively pins the vortex, confining it to a meandering trajectory
over a length scale of order $a$. {\em Right panel:} Length-averaged
vortex speed in response to local SF flow at
$v_s$. Velocities have been scaled by $s^{-3/2}$. For relatively high
$v_s$, the vortex translates at a speed $\sim v_s$, but is trapped for
$v_s<v_c\simeq s^{3/2}$, independent of drag.}
\label{motion}
\end{figure}

\begin{figure*}[t]
\centering
\includegraphics[width=.285\linewidth]{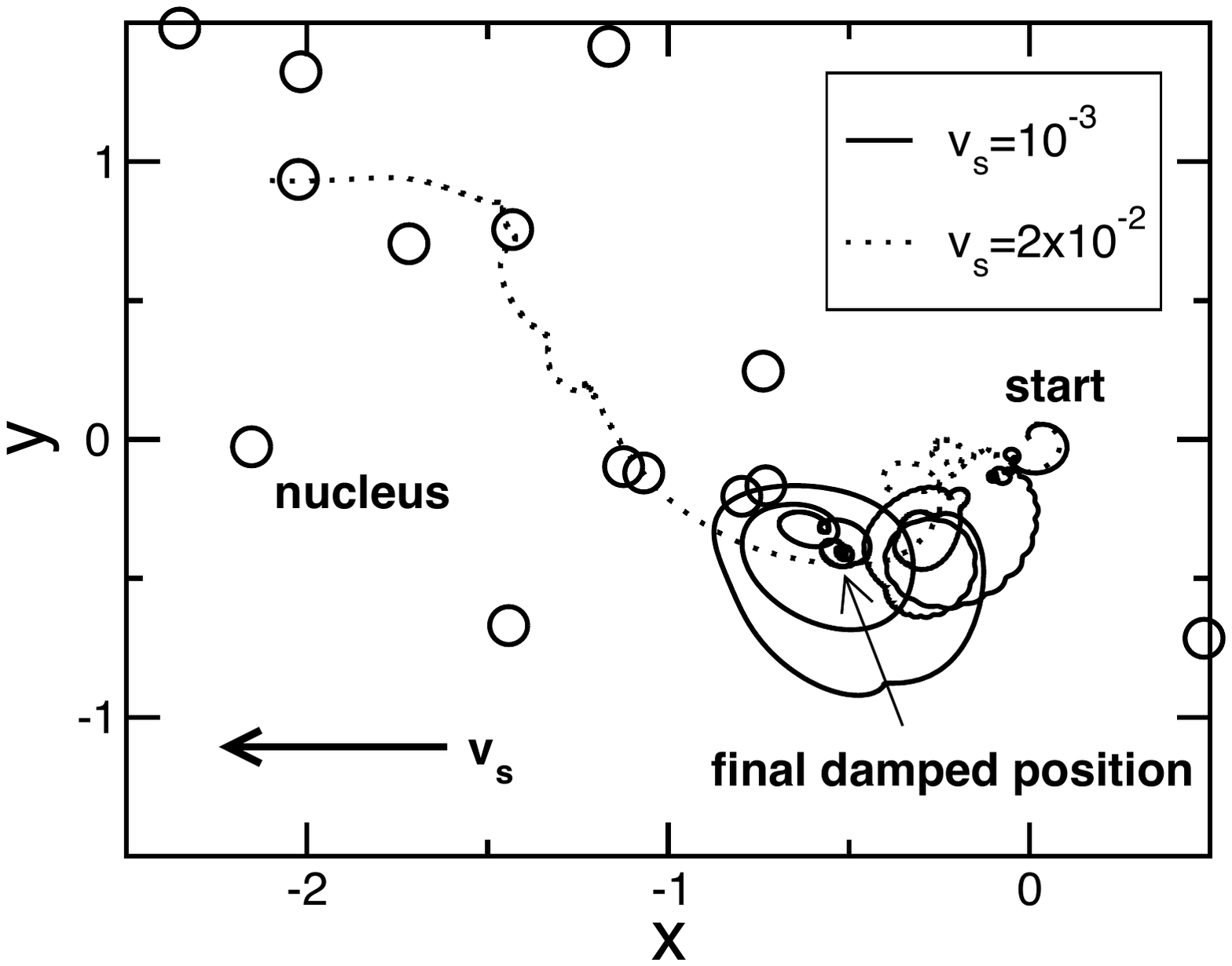} 
\includegraphics[width=.3\linewidth]{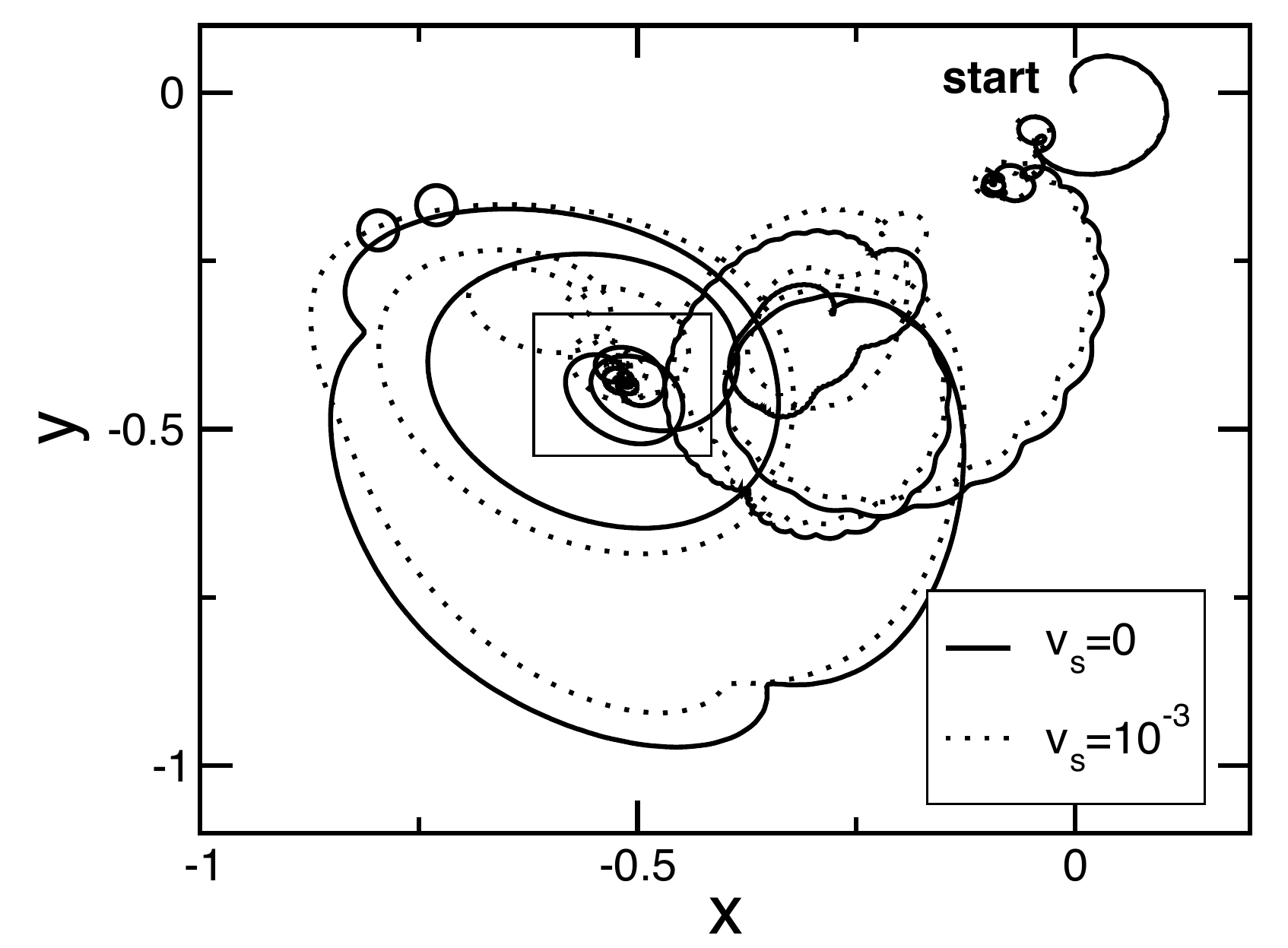} 
\includegraphics[width=.31\linewidth]{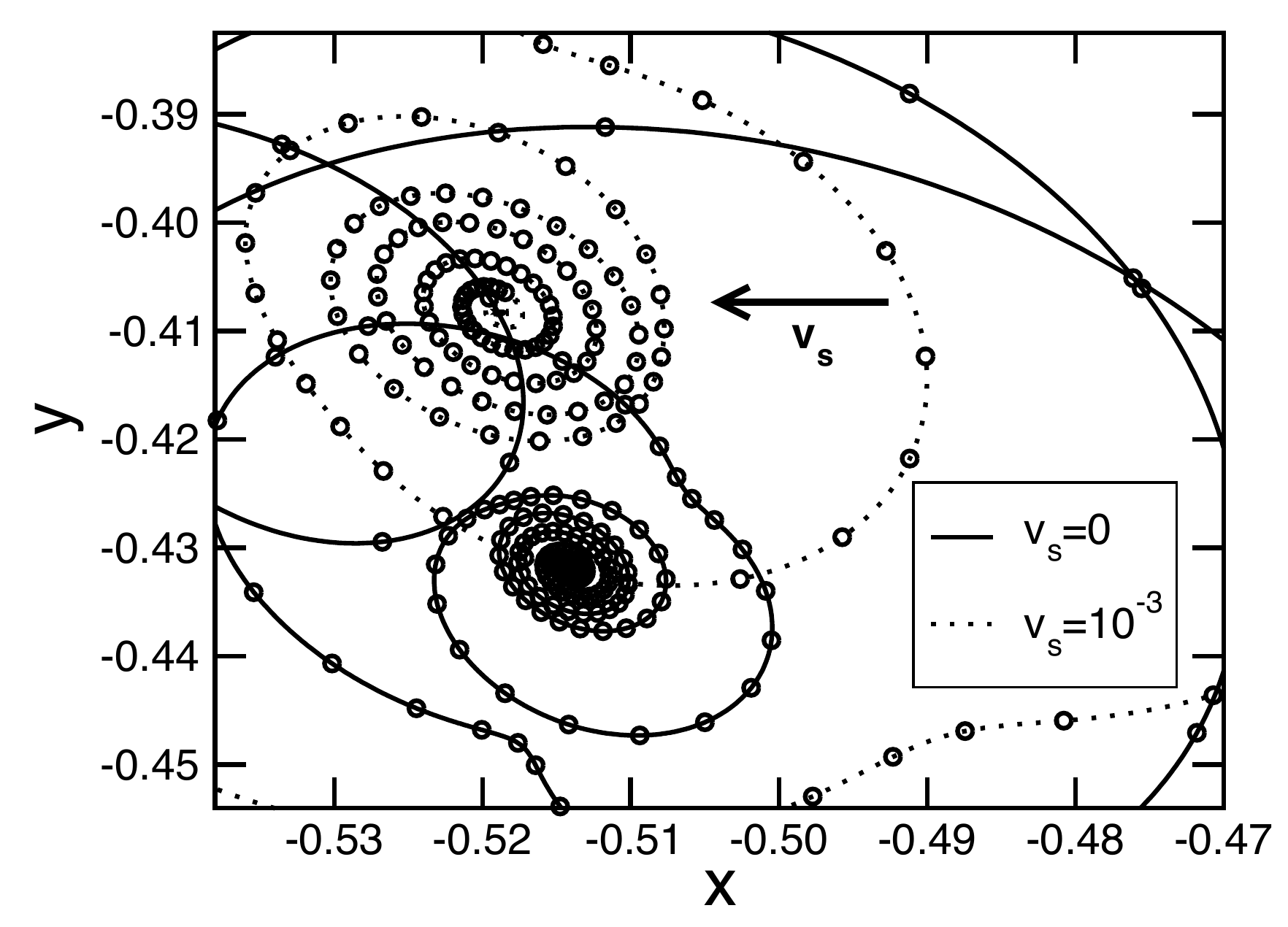}
\caption{Motion of an example vortex segment through one plane of the
lattice for $\eta_r=0.1$. {\em Left panel:} For $v_s>v_c\sim s^{3/2}$, the
vortex translates with the superfluid, with significant changes in
direction as it interacts with nuclei (denoted by circles). For $v_s<v_c$, the
segment damps to a pinned configuration that brings it closer to
nuclei. {\em Middle panel:} Comparison of the damped motion for
$v_s=0$ and $v_s<v_c$, showing that finite $v_s$ has little effect on
the motion in this regime. {\em Right panel:} Detail of the box
region of the middle figure. Circles on the lines indicate points of
equal time. The result of finite $v_s$ is to displace the pinned
vortex slightly in the $y$ direction.  }
\end{figure*}

Comparison of terms in eq. [\ref{eomv}] shows that the dynamics are
essentially determined by the non-dissipative lattice force and $s$ in
the regime $v_s<v_c$. 
To determine the character of the dynamical regimes of vortex motion
and the scaling relations that define them,
eqs. [\ref{eomv}]-[\ref{fl}] were solved numerically. The vortex was 
divided into 50 zones of length $a$, with periodic boundary conditions
applied at the ends, in a random lattice consisting of $10^3$ nuclei
per zone. Fig. \ref{motion} (left panel) shows that the vortex is
trapped by the random potential for finite $v_s$ even for zero
drag. The chief result of the numerical analysis is shown in the right
panel of Fig. 1. Trapping occurs for $v_s$ below a critical velocity
$v_c\simeq s^{3/2}$, {\em independent of drag} in the regime
$\eta_r<<1$. For $v_s>v_c$, the vortex moves at $\vbf\simeq\vsbf$; in
this regime, the drag description of eq. [\ref{drag}] is
adequate. Below $v_c$, however, the vortex velocity is effectively
zero. The drag description fails to show this transition to the pinned
state because it excludes both the non-dissipative lattice force and the
finite vortex tension. The value of the critical velocity $v_c\sim
s^{3/2}$ follows from consideration of a vortex in static equilibrium,
as discussed elsewhere \cite{jones98,lc02}. Only if the vortex had
infinite tension, one of the assumptions implicit in the drag description and
eq. [\ref{drag}], would translatory states exist for any $v_s$.

Dissipative processes will damp the vortex to a stationary, pinned
configuration.  Fig. 2 shows an example segment of the long vortex
moving under drag. For $v_s<v_c$, the trajectory is essentially the
same as for $v_s=0$ since the non-dissipative force dominates the
global Magnus force in this velocity regime. Finite $v_s$ (below $v_c$),
produces only a small displacement of the vortex segment's final,
damped position. Inspection of eq. [\ref{eomv}] suggests that a
dragged vortex will damp to a pinned position over a characteristic
time $\sim (s\eta_r)^{-1}$, and numerical experiments confirm
this. The damping time is $\sim 100$ for the examples shown in
Fig. 2. Fig. 3 shows the initial motion and damping of the vortex in
the $x-z$ plane. The initially-straight vortex forms kinks immediately
in the presence of the random potential, and by $t=10$ the vortex has
been excited to amplitudes of order $a$, the length scale of the
random potential. By $t=10^3$, the vortex has damped to a stationary
pinned configuration with bends over a characteristic length scale of
$\sim 10a$. Because the vortex has large tension, it cannot bend to
intersect every nucleus, but assumes a shape that strikes the best
compromise between the energy gain of being close to a nucleus and the
cost of bending the vortex.

\begin{figure}[b]
\centering
\includegraphics[width=8.cm]{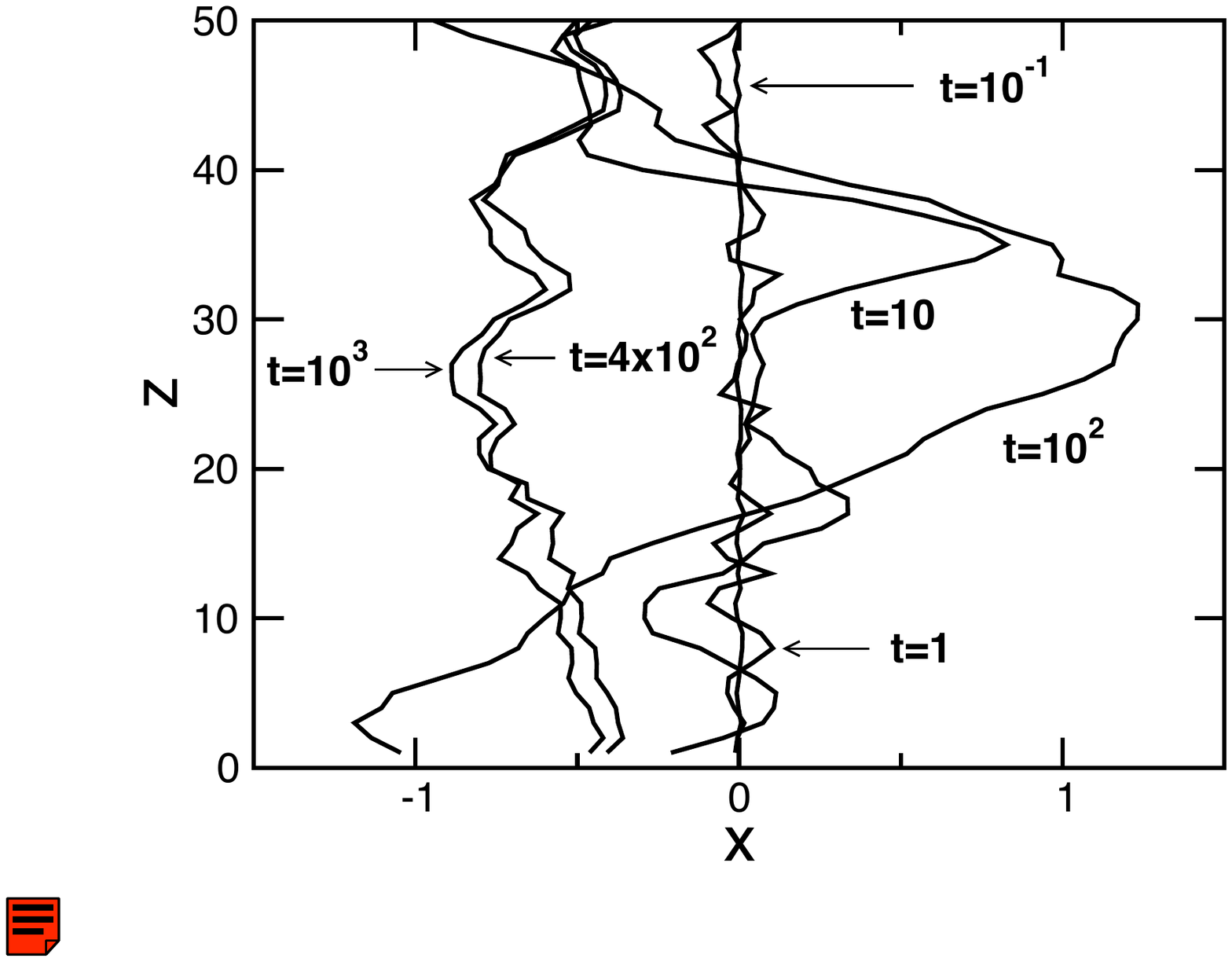} 
\caption{Motion and pinning of the vortex in the $x-z$ plane for the
simulation of Fig. 2 with $\vsbf=0$.}
\end{figure}

Further simulations show that these conclusions are unaffected if the
attractive nuclear potentials are replaced by repulsive ones (though
$v_c$ is slightly reduced for the force of eq. [\ref{force}]). This
result is not surprising, since for $r_p\sim a$ the lattice potential
effectively turns inside out. A vortex segment now damps to a position
that maximizes its distance from the nearest nuclei to the extent
possible against opposing tension forces. For $\eta_r=0$ the vortex is
trapped by the lattice as for the case of attractive nuclei. Since
these results are independent of the value of $\eta_r$ provided
$\eta_r<<1$, they are also independent of the assumption that $\eta_r$
is a constant.

Recent calculations of $E_{vn}$, using the local density approximation
\cite{dp06} and
mean field theory \cite{abbv07} are not in complete agreement. Though
both calculations predict comparable interaction energies ($\sim 1$
MeV) above an inner-crust density of $\sim 10^{13}$ g cm$^{-3}$, they
do not agree on the densities where the interaction is the strongest.
The inevitability of pinning is nevertheless a robust result; it is
insensitive to the exact density range in which $E_{vn}$ is strongest,
the length scale of the potential, or even the sign of the
potential. Within these uncertainties, the critical velocity above
which vortices will translate is
\begin{equation}
v_c\sim
s^{1/2}\frac{F_m}{\rho_s\kappa a} =10^6-10^7 \mbox{ cm s$^{-1}$}.
\label{vc}
\end{equation}
The lattice
will trap the vortex for much smaller values of $E_{vn}$ as well,
though $v_c$ will be lowered according to eq. [\ref{vc}]. 

{\em Conclusions and implications.}---Pinning of vortices to the
lattice of the NS inner crust below a critical value of $v_s$ appears
to be inevitable, whether the interaction is attractive or repulsive
(and provided $E_{vn}$ exceeds the stellar temperature, typically
$\sim 0.01$ MeV). For interactions of order $\sim 1$ MeV per nucleus,
the pinned vortex lattice becomes unstable for $v_s>v_c\sim 10^6-10^7$
cm s$^{-1}$, and vortices will translate approximately with the SF if
the drag is low. For critical velocities this large, the pinned
superfluid can store enough angular momentum to drive the giant
glitches seen in pulsars \cite{lc02}.  If pinning in regions of the
inner crust does occur, however, interpretation of the putative
precession seen in some pulsars becomes problematic; as noted by
Shaham
\cite{shaham77}, if even a small portion of the inner crust vorticity
is tightly coupled to the solid, the star will precess much faster
than the period of $\sim 1$ yr indicated by observations \cite{sls00}.  In the
outer core, where the protons are predicted to form a type-II
superconductor, vortices are expected to pin against a disorganized
system of flux tubes and a similar difficulty arises in explaining
long-period precession \cite{link03}. Anywhere there is pinning,
however, vortices can creep ($v<<v_s$) through thermal activation or
quantum tunneling processes \cite{alpar_etal84,leb93} not considered
here. Vortex creep is incompatible with long-period precession if it
is a high-drag process \cite{swc99,link06}.

\enlargethispage*{10pt}

The core neutron-charge mixture could be unstable to the formation of
SF turbulence in a precessing star \cite{peralta_etal05,gaj08a},
though magnetic stresses might suppress such an instability
\cite{vl08}. [Turbulent instabilities might not occur in the 
single-component SF of the inner crust]. These analyses used the drag
description of vortex motion which does not correctly describe vortex
motion through a potential; it is necessary to include non-dissipative
pinning forces. The mutual friction force of eq. [\ref{mf}] should be
replaced with the {\em total} force per unit volume exerted on the SF
by the vortex array. For a single-component SF, this force is (see,
\eg, \cite{swc99}):
\begin{equation}
{\mathbf F}=\rho_s \left\{\vbf(\vsbf) - \vsbf\right\}\times\omegabf,
\label{ftot}
\end{equation}
where $\vbf(\vsbf)$ is local vortex velocity with respect to the
background, averaged over many vortices, in response to a local flow
$\vsbf$. In the approximations made here, $\vbf=0$ for $v_s<v_c$, and
${\mathbf F}$ reduces to the Magnus force per unit volume on pinned
vortices. At finite $\vsbf$, however, vortices can move slowly through
vortex creep, a process which dissipates energy at a rate
$\rho_s\vbf\cdot (\vsbf\times\omegabf)$ per unit volume; the
dissipative force in the fluid would be small if $\vbf(\vsbf)$ is
nearly orthogonal to $\vsbf\times\omegabf$, while the non-dissipative
force would be, in any case, nearly equal to the Magnus force on
perfectly pinned vortices. It would be interesting to study NS star
modes under the assumption that creep proceeds with little
dissipation, the opposite limit of that studied so far.

\begin{acknowledgments}
I thank I. Wasserman and C. M. Riedel for useful discussions. 
\end{acknowledgments}


\bibliography{references}

\end{document}